\documentclass[10pt,a4paper]{article}

\usepackage[utf8]{inputenc}
\usepackage[T1]{fontenc}
\usepackage[margin=2.0cm]{geometry}
\usepackage[numbered]{bookmark}
\usepackage{mathtools}
\usepackage{ulem}
\usepackage{amssymb}
\usepackage{graphicx}
\usepackage[font=small,labelfont=bf]{caption}
\usepackage{authblk}
\usepackage{cite}
\usepackage{dsfont}
\usepackage{xcolor}

\usepackage[titles]{tocloft}

\numberwithin{equation}{section}

\newcommand{\be}{\begin{equation}}
	\newcommand{\ee}{\end{equation}}
\newcommand{\bea}{\begin{eqnarray}}
	\newcommand{\eea}{\end{eqnarray}}

\newcommand{\nocontentsline}[3]{}
\newcommand{\tocless}[2]{\bgroup\let\addcontentsline=\nocontentsline#1{#2}\egroup}

\title{A note on the Holographic model for color superconductivity in d-dimension without confinement phase}
\author[a,b]{Nguyen Hoang Vu
\footnote{Emails: \texttt{vu@jinr.ru} }}

\affil[a]{\textit{BLTP, JINR,}\textit{141980 Dubna, Moscow region, Russia}} 
\affil[b]{\textit{Institute of Physics, VAST,} \textit{10000, Hanoi, Vietnam}}

\date{}

\begin{document}
	\maketitle

	\begin{abstract}
		In this note, we will generalize the concept of the holography for the color superconductivity (CSC) phase becomes to d-dimension AdS instead of 6d. The dual field theory live in (d-1)-dimension $SU(N_c)$ and have no confinement phase contradiction with the QCD color superconductivity. And we will try to use holographic model with Einstein-Maxwell gravity in d dimension AdS and we study this phase with $N_c\geq 2$. And after, we will discuss the equation of state of the color superconductivity in $d=4$ case without confinement via holography 
	\end{abstract}


\section{Introduction}
In quantum chromodynamics (QCD), the color superconductivity (CSC) is one interesting topic. This is the condensate of two quarks into a Cooper pair, called the diquark, at high chemical potential and low temperature. It similar to how electrons condense in metallic superconductor. However, the strong interaction is the attractive force. Hence, the two quark pair can be created more directly compare with the metallic superconductivity where the the Cooper pair of two electrons is created by the interaction with the phonon. And because this phase occur at high chemical potential, we believe that this phase exist in the inner core of the massive neutron stars \cite{Kazuo 2023}. 

One way to study this phase is use the AdS/CFT correspondence or holography \cite{Maldacena 1997}. In this way, the quantum field theory with strong coupling constant in $d-1$-dimension will be describe by one gravity theory in the $d$-dimension AdS spacetime. The first holographic model for color superconductivity phase is \cite{Basu 2011}, in this paper the authors tried to find the color superconductivity of the quark matter in both the confinement and the deconfiement phase by Einstein-Maxwell gravity. They use the gravitation in the$AdS_6$ which dual to the CSC phase in QCD (the $6d$ RN-AdS black hole is dual to deconfinement phase and the $6d$ AdS soliton dual to the confinemnt phase) both in confinement and deconfinement phase. It is different to other holographic models for QCD where we use the $AdS_5$ to study the other problems in QCD, hologrphic modles for quark gluon plasma \cite{Son2001},\cite{Son2005} and the $AdS_4$ in the papers of metallic superconductivity \cite{Horowitz2008}. This because the color superconductivity phase occur at low temperature (below QCD scale), hence we add one extra dimension which dual to the QCD scale. In more detail, QCD have the confinement phase hence it have the QCD scale. To describe this  scale, we add one compact extra dimension $S^1$ in the boundary which correspond the QCD scale and hence the boundary is $R(1,3)\times S^1$ and the bulk have $6$ dimension ($AdS_6$).

However in \cite{Kazuo 2019} Einstein-Maxwell gravity can only study CSC with $N_c=1$. In more detail, in \cite{Kazuo 2019}, the authors prove that if we only use the Einstein-Maxwell gravity with standard Maxwell interaction, the maximum number of color is only $1.89$. This is the consequent of the BF bound broken \cite{BF1},\cite{BF2}. We can solve this problem by modify gravity \cite{Nam2021} or modify Maxwell interaction \cite{Nam2022}. But here it have one question what happen if we generalized the concept of CSC for $d$-dimension with Einstein-Maxwell gravity and standard Maxwell interaction? Can we study CSC phase with $Nc\geq 2$ only with Einstein-Maxwell and the standard Maxwell interaction?

In our holographic model, we will generalize the notion of the color superconductivity in one arbitrary $SU(N_c)$ instead of the $SU(3)_C$ in QCD. Here, we use the $d$-dimensional Einstein-Maxwell gravity and the holographic for superfluidity/superconductivity to study color superconductivity with gauge symmetry $SU(N_c)$ in the dual quantum field theory. We assume in this gauge field theory, for any value of $N_c$, the confinement phase does not exist, as we will not consider the confined gauge theories in this work. This assumption, while contradicting QCD observed in our universe, is rather expected in general e.g. $\mathcal{N}=4$ SYM, thus does not compromise the generality of our approach. Thus, the scale which analogous to confinement scale does not exist in this paper.And we does not add the compact dimension $S^1$ in this paper (the compact dimension appear only if we consider CSC phase in confined gauge theory via holography). In the second section we will discuss the setup of our model. We will discuss the CSC with $N_c\geq 2$ in the third section. In the forth section, we discuss the equation of state of the color superconductivity in $d=4$ case and near the critical chemical potential case. And finally in the fifth section we resume our result and discuss some things of the next steps.

\section{Model setup}
First of all, we redefine the notion of the color superconductivity in the generalize case. It is the condensate of the Cooper pair consist of two arbitrary fermions in one arbitrary $SU(N_c)$ gauge theory. We assume that, in this $SU(N_c)$ gauge theory the coupling constant is large and the Cooper pair of these two fermions can be created directly like the QCD color superconductivity. Moreover, it occurs in the $d-1$ dimensional flat spacetime. 

In holographic dictionary, the gravity dual of this theory is the gravity in the $d$-dimensional AdS spacetime. And the action for the $d$-dimensional Einstein-Maxwell gravity during the CSC phase transition is given by:
 \begin{equation}
 S=\int d^dx\sqrt{-g}\left(\mathcal{R}+\frac{(d-1)(d-2)}{L^2}-\frac{1}{4}F^2-|(\partial_{\mu}-iqA_{\mu})\psi|^2-m^2|\psi|^2\right) \ ,
 \label{EM_action}
 \end{equation}
where $F_{\mu\nu}=\partial_{\mu}A_{\nu}-\partial_{\nu}A_{\mu}$ and the cosmological constant is determined by $\Lambda=-\frac{(d-1)(d-2)}{2L^2}$. We then set the AdS radius $L=1$ for convenience. Here, the $U(1)$ gauge field $A_{\mu}$ serves as the dual description one current in the $SU(N_c)$ theory that analogous to the baryon number current in the CSC phase of QCD or the electric current in metallic superconductivity. The complex scalar field $\psi$ is dual to the boundary Cooper pair scalar field (here we called the Cooper pair insted of the diquark) operator; specifically, in the holographic model of QCD color superconductivity, it corresponds to the diquark Cooper pair scalar field operator. The $U(1)$ charge $q$ of this scalar field $\psi$ is associated with the thing in the $SU(N_c)$ theory like the baryon number of the diquark in QCD color superconductivity, and its value is given by $q=\frac{2}{N_c}$ in which $N_c$ counts number of color. And analogous in \cite{Nam2022}, we also consider the $SU(N_c)$ as one global symmetry (because the strong coupling $SU(N_c)$ theory in this paper is generalized from the QCD, hence we also have the conditions that we can consider the $SU(3)_C$ as one global symmetry in \cite{Nam2022}).

To further simplify our model from Eq. \eqref{EM_action}, we focus on $s$-wave (or scalar fields) CSC, because the Cooper pair in this case is the scalar fields, in which the vector field $A_\mu$ and complex scalar field $\psi$ follow the ansatz:
 \begin{equation}
     A_{\mu}dx^{\mu}=\phi(r)dt \ , \  \psi=\psi(r) \ ,
\label{ansatz}
 \end{equation}
where the variations are purely radial. The CSC phase emerges from the condensation of the scalar field Cooper pairs, corresponding to the spontaneous breaking of the $SU(N_c)$ symmetry and one $U(1)$ symmetry of the thing in the $SU(N_c)$ gauge theory which analogous to the baryon symmetry $U(1)_B$ in the QCD CSC (called the $U(1)_{B'}$ symmetry). This $U(1)_{B'}$ is dual to the $U(1)$ symmetry in the bulk.  Assuming that the charge is held fixed, the condensation of the scalar field $\psi$ is driven by the chemical potential, analogous to the baryon chemical potential of quarks in QCD color superconductivity. Near the critical chemical potential, the value of the bulk scalar field approaches zero, and under these conditions, the back reaction of the bulk scalar field on the spacetime metric is negligible. Thus, the back reaction of the matter field is primarily contributed by the $U(1)$ gauge field $A_\mu$ alone.

As mentioned above, the confinement phase is not included in this model, and since the CSC phase transition occurs at one critical temperature $T_c$ \cite{Kazuo 2019}, this temperature is associated with the critical chemical potential $\mu_c$ (similar to the QCD CSC phase in the deconfinement phase). Consequently, the spacetime geometry dual to the non confinement theory is described by the Reissner-Nordström (RN) planar black hole solution, with the metric given by the following ansatz:
 \begin{equation}
  \label{black hole solution}
 ds^2=r^2 \Big[-f(r)dt^2+h_{ij}dx^idx^j \Big]+\frac{dr^2}{r^2f(r)}
 \end{equation}
 where $h_{ij}dx^idx^j=dx_1^2+...+dx_{d-2}^2$ is the line element of the $(d-2)$-dimension hypersurface. The event horizon radius $r_+$ satisfies $f(r_+)=0$. In the holographic dictionary, the temperature of the boundary field theory is associated with the Hawking temperature of this $d$-dimensional RN planar AdS black hole, i.e. 
 \begin{equation}
 T=T_H\equiv\frac{r_+^2f'(r_+)}{4\pi} \ .
 \label{Hawking}
 \end{equation}

Using the ansatz Eq. \eqref{ansatz}, we can obtain the classical equations of motion for the temporal component of the vector field $\phi$ and the complex scalar field $\psi$ to be:
  \begin{equation}
  \label{eom}
\begin{split}
\phi''(r)+\frac{d-2}{r}\phi'(r)-\frac{2q^2\psi^2(r)}{r^2f(r)}\phi(r)&=0 \ , 
\\
\psi''(r)+\left[\frac{f'(r)}{f(r)}+\frac{d}{r}\right]\psi'(r)+\frac{1}{r^2f(r)}\left[\frac{q^2\phi^2(r)}{r^2f(r)}-m^2\right]\psi(r)&=0 \ ,
\end{split}
\end{equation}
in which the blackening function $f(r)$ is given by:
 \begin{equation}
 f(r)=1-\left(1+\frac{3\mu^2}{8r_+^2}\right)\left(\frac{r_+}{r}\right)^{d-1}+\frac{3\mu^2r_+^d}{8r^{d+2}}
 \end{equation}
 In $d=6$ case we have $f(r)=1-\left(1+\frac{3\mu^2}{8r^2_+}\right)\left(\frac{r_+}{r}\right)^5+\frac{3\mu^2r_+^6}{8r^8}$ this equivalence to the holographic model for the QCD CSC in deconfinement region \cite{Kazuo 2019}. 

The temperature of physics system is dual to the Hawking temperature and
 \begin{equation}
 T=\frac{r_+^2f'(r_+)}{4\pi}=\frac{1}{4\pi}\left((d-1)r_+-\frac{9\mu^2}{8r_+}\right),
 \end{equation}
 the condition $T\geq 0$ we have the constraint of $\mu$:
$ \frac{\mu^2}{r_+^2}\leq \frac{8(d-1)}{9}$

 Near the boundary $(r\rightarrow\infty)$ we have the form of the matter fields:
 \begin{equation}\label{asymptotic form}
\begin{split}
\phi(r)&= \mu -\frac{\overline{d}}{r^{d-3}}\\
\psi(r)&=\frac{J_C}{r^{\Delta_-}}+\frac{C}{r^{\Delta_+}}
\end{split}
\end{equation}
where $\mu,\overline{d},J_C$ and $C$ are regarded as the chemical potential, charge density, source, and the condensates value (VEV) of the Cooper pair operator dual to $\psi$, like the diquark Cooper pair in QCD, respectively, and we can see that one constraint $d>3$. The conformal dimension $\Delta_{\pm}$ in this case read
\begin{equation}
\Delta_{\pm}=\frac{1}{2}\left((d-1)\pm\sqrt{(d-1)^2+4m^2}\right)
\end{equation}
And the BF bound is
\begin{equation}
\label{BF bound}
    m^2\geq-\frac{(d-1)^2}{4}
\end{equation}
From the BF bound we see that the bulk scalar field is stability if the mass term $m^2\geq -\frac{(d-1)^2}{4}$. We choose $m^2=2-d$ to have $\Delta_-=1$ and we obtain $\Delta_+=d-2$. Hence
\begin{equation}
    \psi(r)=\frac{J_C}{r}+\frac{C}{r^{d-2}}
\end{equation}
And from \cite{Nam2021} the boundary condition at event horizon is given by
\begin{equation}
    \begin{split}
 &\phi(r_+)=0 \ , \\
        &\psi(r_+)=r_+^2\frac{f'(r_+)\psi'(r_+)}{m^2} \ .
    \end{split}
\label{horizon_condition}
\end{equation}

\section{The emerge of color superconductivity with multicolor}
When the chemical potential is greater than critical chemical potential ($\mu>\mu_c$), the Cooper pair condensed state occur. At the critical point, $\psi =0$ hence in near critical chemical potential (it means $\mu>\mu_c$ but near $\mu_c$) the back reaction of the scalar field is negligible. We obtain the bulk configuration is determined by:
 \begin{equation}
 S=\int d^dx\sqrt{-g}\left(\mathcal{R}-2\Lambda-\frac{1}{4}F^2\right)
 \end{equation}
Here, the solution of gauge field in this case is given by
\begin{equation}
    \phi(r)=\mu\left(1-\left(\frac{r_+}{r}\right)^{d-3}\right)
\end{equation}
Because the CSC phase is the condensate, first of all, we must to have one non stability scalar field. After the collapsing of this non stability scalar field, we have the Cooper pair condensate. From the equation of motion for the scalar field, the second equation of \eqref{eom}, we introduce the effective mass:
 \begin{equation}
    m^2_{eff}=m^2-\Delta m^2= m^2-\frac{q^2\phi^2(r)}{r^2f(r)}
 \end{equation}
By the effective mass we rewrite the equation of motion of the scalar field
$  (\square_{RN}-m^2_{eff})\psi_{eff}=0$ where $\square_{RN}=\frac{1}{\sqrt{-g}}\partial_M(\sqrt{-g}g^{MN}\partial_N)$ and the $\psi_{eff}$ is the effective bulk scalar field. To the Cooper pair condensed state appear, we must to have the instability of the bulk scalar field (with the effective mass). And the instability of the bulk fields with the effective mass correspond to the BF bound is broken with the mass square $m^2= m^2_{eff}$. We have a condition
\begin{equation}
\label{BF broken}
    m^2-\frac{q^2\phi^2(r)}{r^2f(r)}<\frac{-(d-1)^2}{4}
\end{equation}
Plug the value of $\phi(r)$ and $f(r)$ and $m^2$ and after some manipulation, we obtain
\begin{equation}
\frac{q^2\hat{\mu}^2z^2\left(1-z^{d-3}\right)^2}{1-\left(1+\frac{3\hat{\mu}^2}{8}\right)z^{d-1}+\frac{3\hat{\mu}^2z^{d+2}}{8}}=q^2F(\hat{\mu},z,d)>\frac{(d-3)^2}{4}
\end{equation}
 with $\hat{\mu}=\frac{\mu}{r_+}$, and $z=\frac{r_+}{r}$
Consider the function: 
\begin{equation}
    F(\hat{\mu},z,d)=\frac{\hat{\mu}^2z^2\left(1-z^{d-3}\right)^2}{1-\left(1+\frac{3\hat{\mu}^2}{8}\right)z^{d-1}+\frac{3\hat{\mu}^2z^{d+2}}{8}} \ ,
\end{equation}
in which we have changed the variable $\mu$ to $\tilde{\mu}$ for simpler constraint condition:
\begin{equation}
\hat{\mu} = \tilde{\mu} \left\{\frac{\left[ 8(d-1)\right]^{1/2}}3\right\} \ \ , \ \     0 < \tilde{\mu} \leq 1 \ .
\end{equation}

 \begin{figure*}[!htbp]
\includegraphics[width=\textwidth]{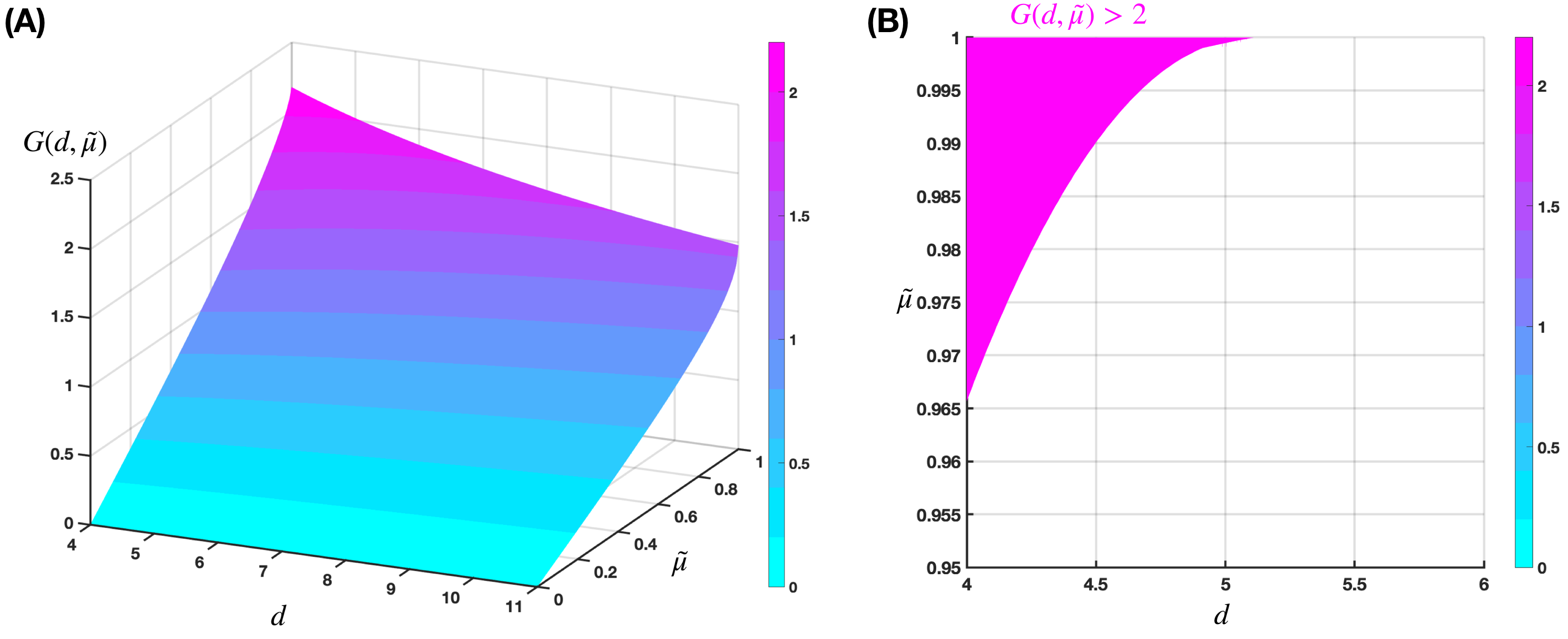}%
\caption{\textbf{Our numerical investigation for $G(d,\tilde{\mu})$ in Eq. \eqref{G_man}.} This calculation was done using MatLab R2023a. \textbf{(A)} The surface function $G(d,\tilde{\mu})$ inside the region of interests i.e. $(d,\tilde{\mu}) 
\in [4,11] \times [0,1]$. \textbf{(B)} We zoom into the small corner where $G(d,\tilde{\mu})>2$ can be realized.}
\label{fig01}
\end{figure*}

This constraint arises because the temperature of our system equals the Hawking temperature of the AdS black hole, which must be greater than or equal to zero. We then define the following two functions:
\begin{equation}
  F_{\max}(d,\tilde{\mu}) = \max_{z\in [0,1]} F(d,\tilde{\mu},z) \ \ , \ \ G(d,\tilde{\mu}) \equiv \frac{4 \left[F_{\max}(d,\tilde{\mu}) \right]^{1/2}}{d-3}   \ ,
  \label{G_man}
\end{equation}
in which the condition for color superconductivity is met when the inequality 
\begin{equation}
    G(d,\tilde{\mu})> N_c
    \label{G_cond}
\end{equation}
holds, according to the Breitenlohner-Freedman (BF) bound \cite{Kazuo 2019}.

Figure \ref{fig01}A presents our numerical investigation of $G(d,\tilde{\mu})$ for $d \in [4,11]$, covering the dimensionality from 4 to 11 dimension supergravity \cite{Natsuume2015} (where fundamental objects are conjectured to be membranes \cite{Vo2024Size} rather than strings \cite{Karliner1988Size}). We find no region where $G(d,\tilde{\mu}) = 3$ and only a small area, shown in Fig. \ref{fig01}B, where $G(d,\tilde{\mu}) > 2$. From the condition Eq. \eqref{G_cond}, our results suggest that color superconductivity for $N_c=2$ can be described via holography using only Einstein-Maxwell gravity, but this is valid only in the case of $d=4$. 


Moreover, we can use this holographic model to show that there is no color superconductivity in $2d$ for all gauge theory (in non confinement case). In fact, the dual gravity theory in this case is $3d$ Einstein -Maxwell gravity. The gauge field solution in $3d$ case is $\phi(r)=0$ and from the condition \eqref{BF broken} we have:
\begin{equation}
2-3-0<-\frac{(3-1)^2}{4}
\end{equation}
The result is $-1<-1$, contradiction. This claim is ensured by the Coleman-Mermin-Wagner's theorem (in $2d$ spacetime, there is no Goldstone boson and hence no superconductivity and color superconductivity and $2d$ spacetime \cite{Natsuume2015}) 

\section{The $d=4$ CSC equation of state}
In this part, we consider the $d=4$ case and study the equation of state of the CSC phase with $d=4$ in the case of non confinement phase. It correspond to the $4d$ RN-AdS black hole. In  $d=4$ case, the blackening function is given by
\begin{equation}\label{blackening function 4d}
f(r)=1-\left(1+\frac{3\mu^2}{8r_+^2}\right)\left(\frac{r+}{r}\right)^3+\frac{3\mu^2r^4_+}{8r^6},
\end{equation}
the Hawking temperature: 
\begin{equation}
T=\frac{1}{4\pi}\left(3r_+-\frac{9\mu^2}{8r_+}\right),
\end{equation}
and hence the chemical potential need to satisfy 
\begin{equation}\label{the mu condition}
\frac{\mu}{r_+}\leq\sqrt{\frac{8}{3}}.
\end{equation}
The equation of motion in this holographic model for $4d$ case is given by
  \begin{equation}
  \label{4d eom}
\begin{split}
\phi''(r)+\frac{2}{r}\phi'(r)-\frac{2q^2\psi^2(r)}{r^2f(r)}\phi(r)&=0 \ , 
\\
\psi''(r)+\left[\frac{f'(r)}{f(r)}+\frac{4}{r}\right]\psi'(r)+\frac{1}{r^2f(r)}\left[\frac{q^2\phi^2(r)}{r^2f(r)}-m^2\right]\psi(r)&=0 \ ,
\end{split}
\end{equation}

Above the critical chemical potential $\mu>\mu_c$, we assume that $J_C=0$ and $C\neq 0$ to ensure that the breaking of the symmetry is spontaneous. To find the equation of state of the $4$ dimension bulk, we need to calculate the Euclidean action. In $d=4$, we have
\begin{equation}
S^E_{d=4}=\int d^4x\sqrt{-g}\mathcal{L}=S^E_{grav}+S^E_{matter}+S^E_{CT}
\end{equation}
where the gravitational part
\begin{equation}\label{gravpart}
S^E_{grav}=\int d^4x\sqrt{-g}\left(\mathcal{R}-2\Lambda\right)
\end{equation}
and the matter part
\begin{equation}
S^E_{matter}=\int d^dx\sqrt{-g}\left(-\frac{1}{4}F^2-|(\partial_{\mu}-iqA_{\mu})\psi|^2-m^2|\psi|^2\right) 
\end{equation}
The gravity part consist of the bulk gravity action $S_{bulkgrav}$ and the Gibbons-Hawking-York $S_{GHY}$ action:
\begin{equation}
 \label{bulk grav}
    \begin{split}
         S_{bulk grav}&=-\int dx^4\sqrt{-g}(R-2\Lambda)=2x^2\beta(4-1)\frac{1}{3}r^3|^{\infty}_{r_+}+\int dx^4\frac{9\mu^2}{4}\frac{r^4_+}{r^6}\\
         &=2X^2\beta r^3|^{\infty}_{r_+}-X^2\beta\frac{3\mu^2}{4}r_+^4\frac{1}{r^3}|^{\infty}_{r_+}=2V_2\beta r^3|^{\infty}_{r_+}+V_2\beta\frac{3\mu^2r_+}{4}
    \end{split}
\end{equation}
 The Gibbons-Hawking-York (G.H.Y) term
 \begin{equation}
     S_{GHY}=-2\int dx^3\sqrt{-h}K
 \end{equation}
  Where $K$ is the extrinsic curvature and $h=det h_{ab}$ with $h_{ab}$ the metric in three dimension boundary, without $r$. The extrinsic curvature is determined by:
\begin{equation}
    K=\frac{1}{2}h^{ab}n^r\partial_rh_{ab}.
\end{equation}  
Hence $h_{ab}=(-r^2f(r),r^2,r^2)$ and $h^{ab}=(-\frac{1}{r^2f(r)},\frac{1}{r^2},\frac{1}{r^2})$, and the normal vector is defined by
\begin{equation}
    n^{\mu}=\frac{1}{\sqrt{g_{rr}}}\left(\frac{\partial}{\partial r}\right)^{\mu}=\frac{\delta_r^{\mu}}{\sqrt{g_{rr}}}
\end{equation}
The non-zero component of the normal vector is 
\begin{equation}
    n^r=r\sqrt{f(r)}.
\end{equation}
 And we obtain
 \begin{equation}
     K=\frac{1}{2}h^{ab}n^r\partial_rh_{ab}=\frac{1}{2}\left(\frac{rf'(r)}{\sqrt{f(r)}}+2\sqrt{f(r)}+4\sqrt{f(r)}\right)
 \end{equation} 
 and
 \begin{equation}
     \sqrt{-h}=\sqrt{-(-r^2f(r).r^2.r^2)}=r^3\sqrt{f(r)}
 \end{equation}
 Hence, the G.H.Y term
 \begin{equation}
 \label{GHY}
     S_{GHY}=\beta V_2(-6r^3f(r)-r^4f'(r))|^\infty
 \end{equation}
 We obtain the gravity part of action is 
 \begin{equation}
 \label{grav action}
 S^E_{grav}=V_2\beta\left(2r^3|^{\infty}_{r_+}+\frac{3\mu^2r_+}{4}-6r^3f(r)|^{\infty}-r^4f'(r)|^{\infty}\right)
 \end{equation}
 
 The matter-part $S^E_{matter}$ is given by
 \begin{equation}
     S^E_{matter}=S^E_{bulk matter}+S^E_{bnd,F}
 \end{equation}
 The bulk matter term is separated into two terms
 \begin{equation}
 S^E_{bulk matter}=S^E_{\psi}+S^E_{\phi}
 \end{equation}
 The first terms is given by (we rewrite $D_{\mu}\psi=(\partial_{\mu}-iqA_{\mu})\psi$)
 \begin{equation}
 \label{equation of Spsi}
 \begin{split}
 S^E_{\psi}&=-\beta V_2\int dr\sqrt{-g}(-|D_{\mu}\psi|^2-m^2|\psi|^2)\\
&=V_2\beta\int dr\sqrt{-g}(g^{rr}\psi'^2+q^2A_0^2\psi^2g^{00}+m^2\psi^2)\\
&=V_2\beta\int dr\sqrt{-g}\left[-\frac{1}{\sqrt{-g}}\partial_r(\sqrt{-g}(g^{rr}\psi'))+q^2A^2_0\psi g^{00}+m^2\psi\right]\psi\\
&+V_2\beta[\sqrt{-g}g^{rr}\psi'\psi]^{\infty}_{r_+}
 \end{split}
 \end{equation}
 The integral part vanishes due to the equation of motion $\frac{1}{\sqrt{-g}}\partial_r(\sqrt{-g}(g^{rr}\psi'))=q^2A^2_0\psi g^{00}+m^2\psi$. And hence:
 \begin{equation}
 S^E_{\psi}=V_2\beta[\sqrt{-g}g^{rr}\psi'\psi]^{\infty}_{r_+}=V_2\beta [r^4f(r)\psi\psi']|^{\infty}_0=0
 \end{equation}
 Because $f(r_+)=0$ and $\psi(r)|_{r\rightarrow\infty}=\frac{C}{r^2}+...$
 For the second term in the $S^E_{bulk matter}$, we have
 \begin{equation}
 \begin{split}
 S^E_{\phi}&=-\beta V_2\int dr\sqrt{-g}\left(-\frac{1}{4}F^2\right)\\
 &=-\beta V_2\int dr\sqrt{-g}\left(-\frac{1}{2}g^{00}g^{rr}\phi'^2\right)\\
&=\beta V_2\left[-\frac{1}{2}\int dr\sqrt{-g}\left[\frac{1}{\sqrt{-g}}\partial_r(\sqrt{-g}g^{00}g^{rr}\phi')\right]\phi+\frac{1}{2}g^{00}g^{rr}\sqrt{-g}\phi\phi'\right]\\
&=\beta V_2\left[-\frac{1}{2}\int dr\sqrt{-g}2q^2g^{00}A_0^2\psi^2+\frac{1}{2}g^{00}g^{rr}\sqrt{-g}\phi\phi'\right]\\
&=\beta V_2[-\int dr\sqrt{-g}q^2g^{00}A_0^2\psi^2+\frac{1}{2}g^{00}g^{rr}\sqrt{-g}\phi\phi']
 \end{split}
 \end{equation}
In $4d$ RN AdS black hole we have
\begin{equation}
S^E_{\phi}=\beta V_2\left[-\int dr \frac{q^2\phi^2\psi^2}{f(r)}+\frac{1}{2}r^2\phi\phi'|^{\infty}_0\right]
\end{equation}
 And the $S^E_{bnd,F}$ is given by
 \begin{equation}
 \begin{split}
S^E_{bnd,F}&=\beta V_2\frac{1}{2}\sqrt{|h|}n_aF^{ab}A_b|^{\infty}\\
&=\beta V_2\frac{1}{2}\sqrt{-h}n_rF^{r0}A_0|^{\infty}\\
&=\beta V_2\frac{1}{2}r^3\sqrt{f(r)}\frac{1}{r\sqrt{f(r)}}\phi'(r)\phi(r)|^{\infty}\\
&=\beta V_2\frac{1}{2}r^2\phi\phi'|^{\infty}
\end{split}
\end{equation}
And we obtain the matter part of the Euclidean action
\begin{equation}
\label{matter action}
S^E_{matter}=\beta V_2\left[-\int dr\frac{q^2\phi^2\psi^2}{f(r)}+r^2\phi\phi'|^{\infty}\right]
\end{equation}
And finally the counter term action of this given by
\begin{equation}
\label{counter term}
S^E_{CT}=2(d-2)\beta V_2\sqrt{-h}|^{\infty}=4\beta V_2r^3\sqrt{f(r)}|^{\infty}
\end{equation}
 From \eqref{grav action}, \eqref{matter action} and \eqref{counter term} we have
 \begin{equation}
 S^E=V_2\beta\left(2r^3|^{\infty}_{r_+}+\frac{3\mu^2r_+}{4}-6r^3f(r)|^{\infty}-r^4f'(r)|^{\infty}-\int dr\frac{q^2\phi^2\psi^2}{f(r)}+r^2\phi\phi'|^{\infty}+4r^3\sqrt{f(r)}|^{\infty}\right)
 \end{equation}
 Replace the blackening function and the expression of the $U(1)$ gauge field component $\phi(r)$ near the critical chemical potential (it is above the critical point but near). We obtain the Eucliean action
 \begin{equation}
 S^E_{d=4}=-\beta V_2\left[r_+^3\left(1-\frac{\mu^2}{8r^2_+}\right)+\int dr\frac{q^2\phi^2\psi^2}{f(r)}\right]
\end{equation}  
We imply that the free energy density of the 4 dimension RN AdS black hole:
\begin{equation}
\Omega_{4dbh}=-r^3_+\left(1-\frac{\mu^2}{8r^2_+}\right)-\int dr\frac{q^2\phi^2\psi^2}{f(r)}
\end{equation}
and the pressure of the color superconducting gas is given by
\begin{equation}
\label{pressure}
p=-\Omega_{4dbh}=r^3_+\left(1-\frac{\mu^2}{8r^2_+}\right)+\int^{\infty}_{r_+} dr\frac{q^2\phi^2\psi^2}{f(r)}
\end{equation}
To obtain the equation of state $p=p(\mu)$, we need to solve the $\psi$ near critical chemical potential. Near critical chemical potential (above but near), we can estimate the $\psi$ by the Sturm-Liouville method. We introduce the new variable
\begin{equation}
z=\frac{r_+}{r},
\end{equation}
with the variable $z$ the gauge field is
\begin{equation}
\phi(z)=\mu(1-z),
\end{equation}
and we have the blackening function for $4d$ bulk ( we set $r_+=1$)
\begin{equation}
f(z)=1-\left(1+\frac{3\mu^2}{8}\right)z^3+\frac{3\mu^2}{8}z^6. 
\end{equation}
The equation of motion for the scalar field $\psi$ is given by
\begin{equation}
\label{eom z variable}
\begin{split}
   &\psi''(z)+\left[\frac{f'(z)}{f(z)}-\frac{2}{z}\right]\psi'(z)+\left[\frac{q^2\phi^2(z)}{f^2(z)}-\frac{m^2}{f(z)z^2}\right]\psi(z)=0
\end{split}
\end{equation}
From the boundary condition of $\psi$, we have the form of $\psi(z)$ as follow:
\begin{equation}
    \psi=Cz^2H(z)
\end{equation}
Where, the function $H(z)$ is the trial function and satisfies the boundary conditions $H(0)=1$ and $H'(0)=0$ (the form of the trial function is chosen). By consider $\mu\gtrapprox\mu_c$, we obtain the equation for $H(z)$:
\begin{equation}
\begin{split}
     &H''(z)+\frac{f'(z)z+2f(z)}{zf(z)}H'(z)+\frac{2[f'(z)z-f(z)]-m^2}{z^2f(z)}H(z)+q^2\mu^2_{c}\frac{(1-z)^2}{f^2(z)}H(z)=0.
\end{split}
\end{equation}
We can rewrite this equation in the form,
\begin{equation}
    H''(z)+p(z)H'(z)+q(z)H(z)+\lambda^2 w(z)\xi^2(z)H(z)=0
\end{equation}
where $\lambda^2=q^2\mu_c^2$ and
\begin{equation}
    \begin{split}
        &p(z)=\frac{f'(z)}{f(z)}+\frac{2}{z}\\
        &q(z)=\frac{2}{z}\left(\frac{f'(z)}{f(z)}-\frac{1}{z}\right)-\frac{m^2}{f(z)z^2}\\
        &=\frac{2[-1-4\left(1+\frac{3\mu_c^2}{8}\right)z^3+\frac{15\mu_c^2}{8}z^6]}{[1-\left(1+\frac{3\mu_c^2}{8}\right)z^3+\frac{3\mu_c^2}{8}z^6]^2}\\
        &-\frac{m^2}{[1-\left(1+\frac{3\mu_c^2}{8}\right)z^3+\frac{3\mu_c^2}{8}z^6]z^2}\\
       & w(z)=\frac{1}{f^2(z)}=\frac{1}{[1-\left(1+\frac{3\mu_c^2}{8}\right)z^3+\frac{3\mu_c^2}{8}z^6]^2}\\
       &\xi^2(z)=(1-z)^2
    \end{split}
\end{equation}
It is written in the form of the Sturm-Liuoville equation.
\begin{equation}
\label{Sturm-Liouville equation}
    [T(z)H'(z)]'-Q(z)H(z)+\lambda^2P(z)H(z)=0
\end{equation}
where
\begin{equation}
\label{S-L coeff}
    \begin{split}
        Q(z)&=-T(z)q(z)\\
        P(z)&=T(z)w(z)\xi^2(z)\\
        T(z)&=e^{\int p(z)dz}=f(z)z^2\\
        &=[1-\left(1+\frac{3\mu_c^2}{8}\right)z^3+\frac{3\mu_c^2}{8}z^6]z^2
    \end{split}
\end{equation}
From the Sturm-Liouville equation, the eigenvalue $\lambda^2$ in \eqref{Sturm-Liouville equation} is obtained by minimizing the following expression:
\begin{equation}
\label{eigenvalue}
    \lambda^2=\frac{\int_0^1 T(z)H'^2(z)dz+\int^1_0 Q(z)H(z)^2dz}{\int^1_0 P(z)H^2(z)dz}.
\end{equation}
We need the $\mu_c<\sqrt{\frac{8}{3}}$ with both $N_c=1$ and $N_c=2$ to satisfy the temperature condition. Hence the trial function $H(z)$ must to have one form such that the right hand side of \eqref{eigenvalue} satisfy the condition: 
\begin{equation}
\label{condition 4d without confinement}
    \begin{split}
        &\text{RHS}_{f(z)=1-2z^3+z^6}<8/3\\
        &\text{RHS}_{f(z)=1-z^3}>0.
    \end{split}
\end{equation}
Where the RHS is the right hand side of the eq.\eqref{eigenvalue}. This condition is the condition of the trial function $H(z)$ to have the CSC phase in $3d$ boundary in the non confinement case in this holographic model. Replace the form of $\phi(z)$, $\psi$ and use the variable $z$ we obtain (with $r_+=1$):
\begin{equation}
p=\left(1-\frac{\mu^2}{8}\right)+\int^1_0 dz\frac{4\mu^2(1-z)^2CH(z)}{1-\left(1+\frac{3\mu^2}{8}\right)z^3+\frac{3\mu^2}{8}z^6},
\end{equation}
with $N_c=1$ and
\begin{equation}
p=\left(1-\frac{\mu^2}{8}\right)+\int^1_0 dz\frac{\mu^2(1-z)^2CH(z)}{1-\left(1+\frac{3\mu^2}{8}\right)z^3+\frac{3\mu^2}{8}z^6}
\end{equation}
with $N_c=2$ case. These are the equations of states of the $3d$ CSC phase with $N_c=1$ and $N_c=2$ in the form $p=p(\mu)$ via holography in $4d$ bulk without confinement case. 

With $N_c=1$, the energy density of CSC phase in $3d$ without confinement is given by
\begin{equation}
\begin{split}
\epsilon &=\mu\frac{\partial p}{\partial\mu}-p\\
&=-1-\frac{\mu^2}{8}+\mu^2\int^1_0 dz\frac{4(1-z)^2CH(z)}{1-\left(1+\frac{3\mu^2}{8}\right)z^3+\frac{3\mu^2}{8}z^6}+3\mu^4C\int^1_0 dz\frac{(1-z^3)z^3(1-z)^2H(z)}{\left[1-\left(1+\frac{3\mu^2}{8}\right)z^3+\frac{3\mu^2}{8}z^6\right]^2}
\end{split}
\end{equation} 
This energy density is positive when $H(z)$ satisfy
\begin{equation}
\mu^2\int^1_0 dz\frac{4(1-z)^2CH(z)}{1-\left(1+\frac{3\mu^2}{8}\right)z^3+\frac{3\mu^2}{8}z^6}+3\mu^4C\int^1_0 dz\frac{(1-z^3)z^3(1-z)^2H(z)}{\left[1-\left(1+\frac{3\mu^2}{8}\right)z^3+\frac{3\mu^2}{8}z^6\right]^2}>1+\frac{\mu^2}{8}.
\end{equation}

And in $N_c=2$ case, the energy density of CSC phase in $3d$ without confinement is given by
\begin{equation}
\begin{split}
\epsilon &=\mu\frac{\partial p}{\partial\mu}-p\\
&=-1-\frac{\mu^2}{8}+\mu^2\int^1_0 dz\frac{(1-z)^2CH(z)}{1-\left(1+\frac{3\mu^2}{8}\right)z^3+\frac{3\mu^2}{8}z^6}+\frac{3}{4}\mu^4C\int^1_0 dz\frac{(1-z^3)z^3(1-z)^2H(z)}{\left[1-\left(1+\frac{3\mu^2}{8}\right)z^3+\frac{3\mu^2}{8}z^6\right]^2}
\end{split}
\end{equation} 
This energy density is positive when $H(z)$ satisfy
\begin{equation}
\mu^2\int^1_0 dz\frac{(1-z)^2CH(z)}{1-\left(1+\frac{3\mu^2}{8}\right)z^3+\frac{3\mu^2}{8}z^6}+\frac{3}{4}\mu^4C\int^1_0 dz\frac{(1-z^3)z^3(1-z)^2H(z)}{\left[1-\left(1+\frac{3\mu^2}{8}\right)z^3+\frac{3\mu^2}{8}z^6\right]^2}>1+\frac{\mu^2}{8}
\end{equation}

\section{Conclusion and discussion}
In non confinement case, if we use only the Einstein-Maxwell gravity with standard Maxwel interaction, we can describe only the color superconductivity with $N_c=2$ in $4d$ bulk via holography. Moreover the trial function $H(z)$ must satisfy the condition \eqref{condition 4d without confinement} to obtain the critical chemical potential of the CSC with $N_c=2$ for $d=4$ bulk spacetime. For $N_c \geq 3$, with the Einstein-Maxwell gravity and the standard Maxwell interaction only, we cannot study the CSC phase for any number of bulk dimensions $d$. Thus, it is necessary to employ modified gravity, such as Einstein-Gauss-Bonnet \cite{Nam2021}, or adjust Maxwell's equations \cite{Nam2022} to study color superconductivity for $N_c \geq 3$ with $d\geq 4$. Moreover, this model also prove that there is not CSC with $2d$. This accord with the Coleman-Mermin-Wagner theorem that show us that in $2d$ have no spontaneous symmetry breaking. 

In this project we only study CSC phase without confinement. We will comeback with the confined gauge theory, like QCD. In that project we will discuss the confinement-deconfinement phase transition in $d$- dimension, we find the critical chemical potential in the confinement-deconfinement in $d$ dimension and we study the CSC in the confinement phase. Moreover we will continue with color superconductivity in magnetic via holography and we will compute the backreaction of the matter fields $\psi$, the holographic entanglement entropy in this model. We hope that we will come back soon.     

\section*{Acknowledgment}
We would like to thank Nam H. Cao for his helpful feedback and insightful discussions. We also acknowledge Trung V. Phan for his assistance in generating Fig. \ref{fig01} and editing this manuscript. Additionally, we are grateful to Dmitry Voskresensky for his valuable discussions on color superconductivity.

\end{document}